%Paper: hep-ph/9310249
%From: taltherr@vxcrna.cern.ch
%Date: Fri, 8 Oct 1993 14:41:01 +0100

%
%  This is a TeX file. It uses also psbox.tex to include figures in the text
%  (the package is available at the Bulletin Board, via "get psbox.tex").
%  The 2 PostScript figures are attached at the end and should be put
%  in 2 different files, banff1.eps and banff2.eps
%
%
\input psbox.tex
\headline={\ifnum\pageno=1\firstheadline\else
\ifodd\pageno\rightheadline \else\leftheadline\fi\fi}
\def\firstheadline{\hfil}
\def\rightheadline{\hfil}
\def\leftheadline{\hfil}
	\footline={\ifnum\pageno=1\firstfootline\else\otherfootline\fi}
\def\firstfootline{\rm\hss\folio\hss}
\def\otherfootline{\hfil}

\font\twelvebf=cmbx10 scaled\magstep 1
\font\twelverm=cmr10 scaled\magstep 1
\font\twelveit=cmti10 scaled\magstep 1

\font\tenbf=cmbx10
\font\tenrm=cmr10
\font\tenit=cmti10

\parindent=1.5pc
\hsize=6.0truein
\vsize=8.5truein
\nopagenumbers
\def\Tr   {{\rm Tr}}
\def\Re   {{\rm Re}}

\def\Ks   {\rlap/K}

\def\Qs   {\rlap/Q}

\def\(    {\left( }    \def\)   {\right) }
\def\[    {\left[}    \def\]   {\right] }
\def\txt #1 {\qquad {\rm #1} \qquad}
\def\lsim{\; \raise0.3ex\hbox{$<$\kern-0.75em\raise-1.1ex\hbox{$\sim$}}\; }
\def\gsim{\; \raise0.3ex\hbox{$>$\kern-0.75em\raise-1.1ex\hbox{$\sim$}}\; }
\def\@versim#1#2{\lower0.2ex\vbox{\baselineskip\z@skip\lineskip\z@skip
  \lineskiplimit\z@\ialign{$\m@th#1\hfil##\hfil$\crcr#2\crcr\sim\crcr}}}

\centerline{\tenbf PLASMON DECAY: FROM QED TO QCD}
\baselineskip=22pt
\vglue 0.8cm
\centerline{\tenrm Tanguy ALTHERR}
\baselineskip=13pt
\centerline{\tenit Theory Division, CERN}
\baselineskip=12pt
\centerline{\tenit CH-1211 Geneva, Switzerland}
\vglue 0.8cm
\centerline{\tenrm ABSTRACT}
\vglue 0.3cm
{\rightskip=3pc
 \leftskip=3pc
 \tenrm\baselineskip=12pt\noindent
Upon using the same theoretical framework, I
describe two interesting decay processes: the electromagnetic plasmon decay
into neutrinos, which can be the dominant cooling mechanism for red giants and
white dwarfs, and the gluonic plasmon decay into quarks, which can be
measured in ultra-relativistic heavy-ion collisions.
\vglue 0.6cm}

%vfil
\twelverm
\baselineskip=14pt
\leftline{\twelvebf 1. Introduction}
\vglue 1pt
Plasmons are collective excitations of bosonic type. Phonons, photons or gluons
build up plasmons in a thermal environment. One important feature of the
plasmon mode is certainly its mass, usually proportional to $gT$, where
$g$ is the coupling constant of the theory and $T$ the temperature. Being
massive, this mode can decay into lighter particles. This phenomenon
cannot happen in vacuum, photons and gluons being strictly massless.

Here, I will describe two different processes: the QED plasmon decay
into neutrino-antineutrino pairs,
 which is of crucial relevance to stellar
cooling, and the QCD plasmon decay into quark-antiquark pairs,
 which could be
measured at future accelerators if a quark-gluon plasma is formed in
heavy-ion collisions.

\vglue 0.3cm
\leftline{\twelvebf 2. Plasmon decay in QED}
\vglue 1pt
The plasmon decay process is one of the dominant cooling mechanisms
 for stars
composed of a degenerate core. It is particularly relevant for the stellar
evolution of white dwarfs and red giants.
In these two systems, the core is composed of degenerate matter, with density
$\rho\simeq 10^6\rm g/cm^3$ and temperature in the range
$T=10^7$-$10^8{\rm K}$.
All electronic levels below the Fermi sphere are occupied.
A typical value of the electron momentum is $p_F=400\ \rm keV$.

The plasmon decay process was first considered a long time ago, by Adams,
Ruderman and Woo$^1$, who were followed by many others$^2$.
The subject has recently been revived when it was realized that the
plasmon dispersion relations that were used did not incorporate the
relativistic effects$^3$. The latest works that compensate for these
effects are due to Itoh et al.$^4$, Braaten and Segel$^5$, and
Haft, Raffelt and Weiss$^6$.

Here, I present a slightly
 different method for calculating the
plasmon decay process$^7$. What is first needed is the effective neutrino
coupling with the electromagnetic field.
Within the Standard Model, one-loop thermal corrections bring in an effective
charge for a neutrino in a medium$^8$:
$$
          \Gamma^\mu = ie_\nu\gamma^\mu L
,\eqno(1)$$
where $L={1\over 2}(1-\gamma_5)$ is the standard left-handed projector and
the effective charge $e_\nu$ is given by
$$
           e_\nu = {2\sqrt{2}\over e} G_F c_V \omega_\beta^2
,\eqno(2)$$
with $c_V={1\over 2} + 2\sin^2\theta_W$.
The neutrino effective charge is matter-dependent through the parameter
$$
        \omega_\beta^2 ={e^2\over 2\pi^2} p_F E_F
          \[ 1 - {1-v_F^2\over 2v_F} \ln{1+v_F\over 1-v_F} \]
,\eqno(3)$$
calculated for the case of  a degenerate electron gas,
where $v_F=p_F/E_F$ is the Fermi velocity.

Because of their weak interactions, the neutrinos emitted from the star are
not thermalized and escape freely from the system. Unlike photons, they can
drain the energy from the core of the star and cool it down more quickly.
Using the cutting rules of Kobes and Semenoff$^9$, the neutrino
(antineutrino) production rate due to transverse photon decay is given by
$$
\eqalign{R_T = {dN_\nu\over d^4x} &= e_\nu^2
\int {d^4Q\over (2\pi)^4} \int {d^4K\over (2\pi)^4} \,
2\pi\delta(K^2)2\pi\delta((Q-K)^2) (n_B (\omega) +\theta(-q^0)) \cr
\times &2\pi\delta(Q^2-\Re\Pi_T(Q))
\Tr \left[ \Ks \gamma_\mu (\Ks + \Qs )\gamma_\nu \right]
{\cal P}^{\mu \nu}_T.\cr}
\eqno(4)$$
In this equation, $K$ and $Q$ are the neutrino and photon four-momenta,
 respectively,
and $n_B$ is the Bose-Einstein distribution function.
The transverse photon projection operator is ${\cal P}^{ij}_T =
-\delta^{ij} + q^i q^j/q^2$, with all other components zero$^{10}$.

After performing the integrations, one arrives at the final analytic
expression
$$
R_T = {e_\nu^2\over 24\pi^3} \int_0^\infty {q^2dq\over \omega} Z_T(Q)
n_B(\omega) Q^2
,\eqno(5)$$
where $Z_T(Q)$ is the Jacobian from the $\delta(Q^2-\Re\Pi_T(Q))$
integration. It has a complicated form$^5$ (numerically,
its value is never far from 1).
It must be understood that in Eq.~(5)  $\omega$ and $q$ are related
by a dispersion relation:
$$
\omega^2 - q^2 = {3\over 2} \omega_0^2 \[ {\omega^2\over v_F^2 q^2}
   + {1\over 2v_F} {\omega \over q} \( 1- {\omega^2\over v_F^2 q^2} \)
  \ln {\omega  + v_F q \over  \omega - v_F q }\]
,\eqno(6)$$
where $\omega_0$ is the plasmon frequency given by
$$
\omega_0^2 = {e^2\over 3\pi^2} {p_F^3\over E_F}
.\eqno(7)$$
Similar relations can be derived for the longitudinal case.
These dispersion relations were first derived by Jancovici$^{11}$,
 but had a much
more complicated form. It was realized later that simplified expressions
could be used in a much wider regime$^{5,12}$.
Note also that for the two stellar systems of physical interest, namely the
white dwarfs and the red giants, one can completely
neglect the temperature effects in the plasmon dispersion relations.

The expression for the longitudinal emissivity is exactly the same as in
Eq.~(5), appart from an overall factor $1/2$.
 However, one should not forget
that the dispersion relations quite differ.

To be honest, Eq.~(5) is only valid when the plasmon is sufficiently
close to the light-cone. The neutrino effective charge is in fact a
charge radius$^{1-7}$:
$$
           e_\nu = {2\sqrt{2}\over e} G_F c_V Q^2
,\eqno(8)$$
and one has $Q^2=\omega_\beta^2$ when $\omega,q\gg\omega_0$, as it
should.

Finally, the energy loss rate due to $\nu\bar\nu$ emission is simply obtained
by multiplying Eq.~(5) by the photon energy $\omega$ under the
integral.
$$\psboxto(10.cm;10.cm){banff1.eps}$$

Using the non-relativistic dispersion relations, one easily recovers the
old standard result$^1$ (taking $\sin^2\theta_W=1/2$)
$$
\epsilon_T^{NR} = {G_F^2\over 48\pi^4\alpha} \omega_0^6
\int_{\omega_0}^\infty d\omega {\omega\sqrt{\omega^2-\omega_0^2} \over
                                e^{\beta\omega}-1 }
.\eqno(9)$$
The effect of the additional cooling mechanism due to
plasmon decay is shown in Fig.~1. At the early stages of the white dwarf
evolution, the
neutrino luminosity can be five times more important than the photon
luminosity. It can be shown that the plasmon decay dominates over other
competing neutrino-emitting processes$^6$.

\vglue 0.3cm
\leftline{\twelvebf 3. Plasmon decay in QCD}
\vglue 1pt

The possibility that ultra-relativistic nuclear collisions
may create a quark-gluon plasma is extremely interesting.
Particularly relevant to this problem are the different time scales
which are involved in these collisions: the thermalization time,
the chemical equilibration time, etc$\ldots$.
Recent work$^{13}$ has shown that, while the hot matter should
thermalize in about $0.3\ \rm fm/c$, it may be largely a gluon plasma (GP),
with few quarks.  Chemical equilibration is expected to take much
longer than thermal equilibration, if it occurs at all.

Let me consider the idealized situation of a GP or QGP in thermal
equilibrium.  If such a plasma does exist, the massless gluons evolve
into quasi-particles with effective masses of order $gT$.
Being massive, these quasi-gluons decay into $q\bar q$ pairs. This
situation is then  very similar to the one studied previously.

The plasma frequency (or plasmon mass) in a QCD plasma is given by$^{10}$
$$
       \omega_0^2 = \left( N+ {N_f\over 2} \right) {g^2 T^2\over 9},
\eqno(9)$$
for $SU(N)$ gauge theory at temperature $T$, where $g$ is the strong
coupling constant and $N_f$ is the number of massless fermion flavours.
Thermal effects in QGP and GP are thus identical, except that $N_f=0$ for
GP.

The gluon decay is an obvious mechanism for the production of $q\bar q$
pairs. It has been overlooked in the past literature$^{14}$.
The starting equation for the rate is exactly the same as in
Eq.~(4), except for the coupling.
However, unlike the photon, the gluon has an anomalously large damping
rate$^{15}$:
$$
   \gamma_g = 3\alpha_S T \ln{\omega_0\over m_{mag}} + O(\alpha_S)
,\eqno(10)$$
where $m_{mag}=O(g^2T)$ is the magnetic mass.
Therefore, instead of having a $\delta(Q^2-\Re\Pi(Q))$ for the gluon
propagator, one has a Lorentzian with width $\gamma_g$.

When this effect is taken into account, one finds for the transverse gluon
decay (and for massless quarks)$^{16}$:
$$
R^T_{g\to q\bar q} = {2g^2\gamma_g\over 9\pi^4}
\int_0^\infty d\omega \ \omega^2\ n_B (\omega) \left\{ \ln
{64\omega^4\over 9\omega_0^4  + 16 \gamma_g^2 \omega^2}
+ {3 \omega_0^2\over 2 \gamma_g \omega} \left( \arctan
{3 \omega_0^2\over 4 \gamma_g \omega} + \arctan{2\omega\over \gamma_g}
\right) - 4 \right\},
\eqno(11)$$
where terms of higher order than $g^4$ have been dropped.
The limit $\gamma_g\to 0$ reproduces the result using a gluon
propagator without a finite damping rate (see Eq.~(5)).

Together with the longitudinal contribution, the final result is
$$
R_{g\to q\bar q}
        = {2\zeta(3) \over \pi^3} \alpha_S^2
         \left( \ln{1\over \alpha_S} \right)^2 T^4
        + {\cal O} \left(\alpha_S^2 \ln{1\over \alpha_S} T^4\right).
\eqno(12)$$
Notice that using a bare gluon propagator, $\delta(Q^2)$, would lead to
a vanishing result.  With just the hard thermal mass,
$\delta(Q^2-3\omega_0^2/2)$, the rate is of order $g^4T^4$, with no
logarithmic dependence. Taking into account the anomalously large
damping rate $\gamma_g$ shifts the gluon on-shellness $Q^2\sim \omega_0^2$
by a logarithmic correction. The additional logarithm has a kinematic
origin and comes from the pole of the gluon propagator that is
almost on shell.

The gluon decay process is the leading contribution in the
perturbative expansion, i.e. when $g\to 0$. Diagrams that were calculated
in previous works, as $gg\to q\bar q$ and $q\bar q\to q\bar q$, are
subleading compared to this process (although just by a log).

Even though, numerical results show that chemical equilibration will
proceed very slowly, with $\tau_q >10\ T^{-1}$, even at $g=3$, for
the very light quarks$^{16}$.
$$\psboxto(10.cm;10.cm){banff2.eps}$$

The picture is a bit changed when massive quarks are considered.
When $M\gg T$, one gets for the rates$^{17}$
$$\eqalign{
    R_{g\to Q\bar Q} &= {2\alpha_S\over \pi^2} \gamma_g T^3 e^{-2M/T}  \cr
    R_{gg\to Q\bar Q} &= {7\alpha_S^2\over 6\pi^2} M T^3 e^{-2M/T}  \cr
    R_{q\bar q\to Q\bar Q} &= {\alpha_S^2\over \pi^2} M T^3 e^{-2M/T}
.\cr}\eqno(13)$$
These rates are plotted in Fig.~2. The gluon decay dominates for light
($M<2T$) quarks. Certainly, one should be able to measure the gluon damping
rate, and therefore the magnetic mass, by looking at the strange or
charm quark production in heavy-ion collisions.

\vglue 0.6cm
\leftline{\twelvebf 4. Conclusion}
\vglue 0.4cm
In the two examples I have given, Thermal Field Theory has proved its power
and its usefulness. In the first case, the analytic expressions are much
simpler to manipulate than those obtained by using Kinetic Theory.
This complexity may have been the reason why relativistic effects were
overlooked for 30 years. The second
example is an even better one, of the
confusion that can arise by using the
Kinetic Theory. It is very encouraging to see that the wonderful structure
of hot QCD can be tested in relativistic heavy-ion collisions. This is
especially true for the gluon damping rate, to which so much theoretical
work has been devoted.

\vglue 0.6cm
\leftline{\twelvebf Acknowledgements}
\vglue 0.4cm
The QED part of this work was done with P.~Salati and the
QCD part with D.~Seibert. I would like to thank them both for their helpful
collaboration.

\vglue 0.6cm
\leftline{\twelvebf References}
\vglue 0.4cm
\itemitem{1.} J.~B.~Adams, M.~A.~Ruderman and C.-H.~Woo, {\twelveit Phys.
Rev.} {\twelvebf 129} (1963) 1383.

\itemitem{2.} M.~H.~Zaidi, {\twelveit Nuovo Cim.} {\twelvebf 40A} (1965)
502; G.~Beaudet, V.~Petrosian and E.~E.~Salpeter, {\twelveit Ap. J.}
{\twelvebf 150} (1967) 979; D.~A.~Dicus, {\twelveit Phys. Rev.} {\twelvebf
D6} (1972) 941.

\itemitem{3.} E.~Braaten, {\twelveit Phys. Rev. Lett.} {\twelvebf 66} (1991)
1655.

\itemitem{4.} N.~Itoh, H.~Mutoh, A.~Hikita and Y.~Kohyama, {\twelveit Ap. J.}
{\twelvebf 395} (1992) 622.

\itemitem{5.} E.~Braaten and D.~Segel, {\twelveit Phys. Rev.} {\twelvebf D47}
(1993) 1478.

\itemitem{6.} M.~Haft, G.~Raffelt and A.~Weiss, to appear in
{\twelveit Ap. J.}

\itemitem{7.} T.~Altherr and P.~Salati, work in preparation.

\itemitem{8.} T.~Altherr and K.~Kainulainen, {\twelveit Phys. Lett.}
{\twelvebf B262} (1991) 79.

\itemitem{9.} R.~L.~Kobes and G.~W.~Semenoff, {\twelveit Nucl. Phys.}
         {\twelvebf B260} (1985) 714 and {\twelvebf B272} (1986) 329.

\itemitem{10.} H.~A.~Weldon, {\twelveit Phys. Rev.} {\twelvebf D26} (1982)
1394.

\itemitem{11.} B.~Jancovici, {\twelveit Nuovo Cim.} {\twelvebf 25} (1962) 428
{}.

\itemitem{12.} T.~Altherr, E.~Petitgirard and T.~del Rio Gaztelurrutia,
              {\twelveit Astropart. Phys.} {\twelvebf 1} (1993) 289.

\itemitem{13.} K.~Geiger, {\twelveit Phys. Rev.} {\twelvebf D46} (1992) 4965;
           E.~Shuryak, {\twelveit Phys. Rev. Lett.} {\twelvebf 68} (1992) 3270.

\itemitem{14.} J.~Rafelski and B.~M\"uller, {\twelveit Phys. Rev. Lett.}
{\twelvebf 48} (1982) 1066; T.~Matsui, B.~Svetitsky and L.~D.~McLerran,
{\twelveit Phys. Rev.} {\twelvebf D34} (1986) 783;
A.~Shor, {\twelveit Phys. Lett.} {\twelvebf B215} (1988) 375.

\itemitem{15.}R.~D.~Pisarski, {\twelveit Phys. Rev. Lett.} {\twelvebf 63}
(1989) 1129 and {\twelveit Phys. Rev.} {\twelvebf D47} (1993) 5589.

\itemitem{16.}T.~Altherr and D.~Seibert, {\twelveit Phys. Lett.} {\twelvebf
B313} (1993) 149.

\itemitem{17.}T.~Altherr and D.~Seibert, CERN preprint CERN-TH-7038/93.
\bye